\documentclass[aip,amsmath,amssymb, reprint]{revtex4-1}
\usepackage{xcolor}
\usepackage{graphicx}
\usepackage{dcolumn}
\usepackage{bm}
\usepackage{soul}
\usepackage[utf8]{inputenc}
\usepackage[T1]{fontenc}
\usepackage{mathptmx}
\usepackage{float}
\usepackage{lipsum}
\usepackage{caption}
\usepackage{subcaption}

\begin{document}

\preprint{AIP/123-QED}

\title{Dynamical anomalies and structural features of  Active Brownian Particles characterised by two repulsive length scales}


\author{Jos\'e Mart\'in-Roca}
\affiliation{ Dep. Est. de la Materia, F\'isica T\'ermica y Electr\'onica, Universidad Complutense de Madrid, 28040 Madrid, Spain}

 \author{Ra\'ul Martinez}
\affiliation{ Dep. Est. de la Materia, F\'isica T\'ermica y Electr\'onica, Universidad Complutense de Madrid, 28040 Madrid, Spain}
\affiliation{Dep. de F\'isica Te\'orica de la Materia Condensada, Facultad de Ciencias, Universidad Aut\'onoma de Madrid, 28049 Madrid, Spain}

 \author{Fernando Martínez-Pedrero}
\affiliation{ Dep. Qui\'mica-Fi\'sica, Universidad Complutense de Madrid, Avda. Complutense s/n, Madrid, 28040, Spain.}

 \author{Jorge Ram{\'i}rez}
\affiliation{Dep. de Ingenier{\'i}a Qu{\'i}mica, ETSI Industriales, Universidad Polit{\'e}cnica de Madrid, 28006 Madrid, Spain.}

 \author{Chantal Valeriani}
\affiliation{ 
Dep. de Est. de la Materia, F\'isica T\'ermica y Electr\'onica, Universidad Complutense de Madrid, 28040 Madrid, Spain
}
\affiliation{ 
GISC - Grupo Interdisciplinar de Sistemas Complejos 28040 Madrid, Spain
}

\begin{abstract}
In this work we study a two-dimensional system composed by Active Brownian Particles (ABPs) interacting via a repulsive potential with two-length-scales, a soft shell and a hard-core. Depending on the ratio between the strength of the soft shell barrier and the activity, we find two regimes: If this ratio is much larger or smaller than 1,  the observed behaviour is comparable with ABPs interacting via a single length-scale potential. If this ratio is similar to 1, the two length-scales are relevant for both structure and dynamical properties. 
On the structural side,  when the system exhibits a motility induced phase separation, the dense phase is characterised by new and more complex structures compared with the hexatic phase observed in single length-scale systems.
{\color{black} From the dynamic analysis we find, to our knowledge, the first manifestation of a dynamic heterogeneity in active particles, reminiscent of the glassy dynamics widely studied in passive colloids.}

\end{abstract}

\maketitle

\section{Introduction}

Active particle suspensions are out-of-equilibrium systems in which particles dissipate energy and transform it into motion\cite{1,2}. A characteristic feature of active particles is collective motion, observed both in active living systems \cite{3} (such as bacteria colonies or flock of birds) as well as in synthetic active particles \cite{4,5}. One of the simplest numerical model introduced to  better understand the physics of active particles is  the so-called Active Brownian Particles (ABPs). ABPs follow the equations of motion of Brownian dynamics, modified to allow self-propulsion and gradual change of the direction of particle motion by rotational diffusion. When ABPs interact repulsively at relatively high density and activity, particles tend to exhibit a process called Motility Induced Phase Separation (MIPS), despite the absence of explicit attractive forces. MIPS \cite{6,7,8,9,10,11,omar2021phase}  is promoted by the greater persistence length supported by the activity, as compared to that sustained by the thermal rotational diffusion. It has recently been shown\cite{13,14,15,16,17,JOSE} that the steepness of the repulsion plays a role in the establishment of MIPS phase boundaries. 

Even though the MIPS dense phase is an assembly of geometrically frustrated particles, it does not display any glassy feature, such as a dynamical slowing down  \cite{paoluzzi2021does}.

A glass transition has been encountered in  driven granular media, active and living matter \cite{berthier2013non}.
Even though  important features of glassy dynamics are insensitive to details, the location of
the transition depends on the specifics of the driving mechanisms. 
When increasing the packing fraction, even a suspension of self-propelled repulsive spheres undergoes a dynamic slowing down. However, differently from a passive system, the glass transition can be shifted to higher packing fractions upon increasing the activity \cite{ni2013pushing}.

\textcolor{black}{An active glass is fundamentally different from an equilibrium
glassy system. Activity does not simply
fluidize the system \cite{paul2021dynamic}, but  increases the 
dynamic heterogeneities, even for the same relaxation times.   
Glass theories such as the random first-order transition (RFOT) theory have been extended to a dense assembly of selfpropelled
particles\cite{nandi2018random}, showing that the behavior of the active glass is strongly influenced by the microscopic details of activity.}
\textcolor{black}{The authors of Ref.\cite{mandal2020extreme} studied the dynamics of a dense binary Lennard Jones mixture of active particles (considering inertia). They  demonstrated that tuning the persistence
time,  the system underwent a crossover between a glassy   
(characterised by  density relaxation) and a jamming behaviour.} 
Only very recently, the glass transition has been studied in a suspension of  complex active particles: a  
dense solution of circular active polymers \cite{smrek2020active}. In this system, the interplay between  activity and  topology of the polymers
generated a unique glassy state.


To the best of our knowledge, investigations on active Brownian particles to date have focused on repulsive interactions with one characteristic length-scale, i.e., excluded-volume interactions.
When dealing with equilibrium (passive) systems, it has been shown that it is possible to generate water-like behaviour (including its anomalies) when particles interact through isotropic potentials with two characteristic length scales: a hard-core (short-range) repulsion combined with soft-shell (long-range)  repulsion. \cite{17-a,19,23,24,26,27,28,31,32,33,34,35,Angelo,paul2021dynamic}\textcolor{black}{\cite{franzese2007differences}}.
This shoulder-type interaction potential has been also used to simulate equilibrium suspensions of block copolymers or colloidal spheres decorated with filaments on their surface \cite{Schultz}.
These systems are characterised by a very rich phase behavior, with various ordered structures, in two \cite{malescio2004stripe, gribova2009waterlike}\textcolor{black}{\cite{caprini2018cluster}} or  three dimensions, which is in contrast with the phase behaviour observed in 
a suspension of particles interacting via a single repulsive core.
Malescio and Pellicane \cite{malescio2004stripe} 
demonstrated that in a two-dimensional suspension of colloids interacting
via a shoulder-type potential, at densities where 
hard-and-soft core radii compete with each other, the decrease in 
temperature induces a transition from a disordered state to
an orientationally ordered phase characterized by stripe patterns.
In three dimensional suspensions, a surprisingly complex phase behaviour has also been established \cite{gribova2009waterlike}.
The authors
detected density and diffusion anomalies similar to those in water and concluded
that the anomalies disappeared as the width of the repulsive
step increased, shifting to the regions inside the crystalline
phase  in the vicinity of the maximum of the melting line.


These anomalous features  appeared  not only in complex liquids
\cite{yan2008correspondence,gribova2009waterlike,barraz2009thermodynamic}.
but also in silica  \cite{Horbach_2008}
and  in colloidal systems \cite{foffi2003structural}.
Considering passive colloidal particles interacting via two repulsive length scales, the authors of Ref.\cite{sperl2010disconnected}
demonstrated that  such systems presented a 
logarithmic decay of the density autocorrelation function and a subdiffusive regime in the mean-square displacement \cite{gnan2014multiple}.
Such anomalous dynamics had also been detected in the reentrant glass behaviour that emerges in a system  of passive  hard spheres with short-range attraction \cite{pham2002multiple}, caused by  the existence of two  different glassy states:  one dominated by repulsion (with structural arrest due to caging) and the other dominated by attraction (with structural arrest due to bonding).
Thus, the anomalous dynamics was a  sign of competing glass transitions in core-softened systems and in repulsive systems with two competing length scales.


In the present work we study a two-dimensional suspension of ABPs interacting with a repulsive shoulder-type potential, to unravel how  two length scales  affect both the structural behaviour or the dynamics of the suspension.



\begin{figure*}
    \centering
    \includegraphics[width=1.0\textwidth]{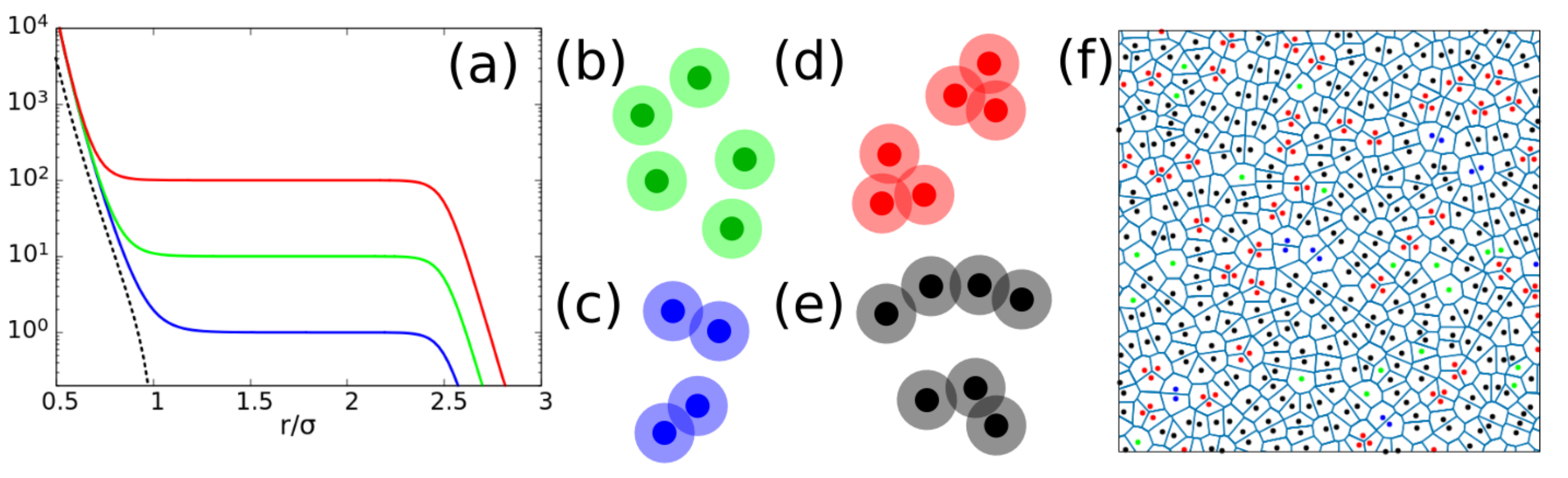}
    \caption{  (a) Semilogarithmic representation of the repulsive potential $V(r)$ vs $r/\sigma$ (see Eq. \eqref{eq:shoulder_potential}), for different values of the shoulder height: $\epsilon_s=1$ (blue), $\epsilon_s=10$ (green) and $\epsilon_s=100$ (red). The remaining parameters adopt the values indicated in the text. The Weeks-Chandler-Andersen (WCA) potential is shown (black dash line) for comparison purposes. (b-e) Examples of different local structures observed in our simulations: monomers (b, green), dimers (c, blue), trimers (d, red) and chains (e, black). (f) Voronoi tessellation of a steady-state snapshot taken from a simulation with Pe=25 and $\epsilon_s=10$. Each of the particles that make up the local structures is represented by the corresponding color, as described above.}
    \label{fig:phase}
\end{figure*}

\section{Simulation  Details}

\subsection{Numerical  Details}
The system consists of $N$ Active Brownian Particles (ABPs) interacting through a hard core potential surrounded by a soft corona \cite{malescio2004stripe}. Particles are located in a two-dimensional box of area $A=L \text{x} L$ with periodic boundary conditions. The equations of motion for the position $\vec{r}_i$ and orientation $\theta_i$ of the $i$-th active particle can be written as
\begin{align}
\label{eq:motion}
& \dot{\vec{r}}_i = \frac{D_t}{k_B T} \left( - \sum_{j\neq i} \nabla V(r_{ij}) + |F_a |\, \vec{n}_i \right) + \sqrt{2D_t} \, \vec{\xi}_i, \\
& \dot{\theta}_i = \sqrt{2D_r}\, \xi_{i,\theta},
\end{align}
where $V(r_{ij})$ is the inter-particle pair potential, $k_B$ is the Boltzmann constant, $T$  the absolute temperature, the components of $\vec{\xi}_i$ and $\xi_{i,\theta}$ are white noise with zero mean and correlations $\langle\xi^{\alpha}_i(t)\xi^{\beta}_j(t')\rangle = \delta_{ij} \delta_{\alpha\beta} \delta(t-t')$, being $\alpha,\beta=x,y$; $F_a$ is a constant self-propelling force acting along the vector $\vec{n}_i$, which forms an angle $\theta_i$ with the positive $x$-axis, and $D_t$ and $D_r$ are the traslational and rotational diffusion coefficients, respectively. For spherical particles at equilibrium, the relation between both coefficients obeys the Stokes-Einstein relation \cite{StokesEinstein}: $D_r = 3D_t/\sigma^2$, where $\sigma$ is the hard core diameter. The inter-particle potential is characterized by two different length scales: a repulsive hard core and a soft repulsive shoulder, given by the expression\cite{gribova2009waterlike}:

\begin{equation}
    V(r)=\epsilon \left( \frac{\sigma}{r} \right)^n + \frac{1}{2} \, \varepsilon_s \, \left\{ 1-\tanh{\left[ k_0 \, \left( r- \sigma_s \right) \right]}\right\},
    \label{eq:shoulder_potential}
\end{equation}
Here, $\epsilon$ is the energy related to the hard core, $\epsilon_s$ and $\sigma_s$ are the height and width of the repulsive shoulder, respectively, $n$ affects the stiffness of the repulsive core and $k_0$ determines the steepness of the shoulder decay (Figure \ref{fig:phase}.a.).  Throughout this work, we have chosen to establish the following parameters: $n=14$ and $k_0 =10/\sigma$ as in ref\cite{gribova2009waterlike}, and $\sigma_s=2.5 \sigma$. In order to study the time evolution of the system we have used the open source code {\color{black} LAMMPS\cite{LAMMPS} }; when dealing with the {\color{black} $N=2^{14}$} systems, however we have used a modified version of UAMMD\cite{UAMMD} (Universally Adaptable Multiscale Molecular Dynamics). Each simulation has been reiterated until the system attains steady state, which we assume arrives when there are no further significant changes in potential energy and overall phase behaviour. The level of activity is measured via the Peclet number, defined as
\begin{equation}
    \text{Pe} \equiv \frac{3\,v_p \, \tau_r }{\sigma} = \frac{3\, |F_a| \, D_t}{ k_B \, T\, D_r \, \sigma},
    \label{eq:peclet}
\end{equation} 
where $v_p=|F_a|D_t/k_BT$ is the velocity propulsion of  the active particles
and $\tau_r=1/D_r$ the reorentation time.Throughout the manuscript, we have varied the Peclet number by changing $D_r$, keeping all other parameters fixed. Other ways to modify Pe (for example, by changing magnitude of the active force) yield slightly different phase behaviour\cite{JOSE}. All quantities are expressed in reduced Lennard-Jones units, with lengths, times and energies given in  terms of $\sigma=\tau =\epsilon=1$, where $\tau=\sqrt{m\sigma^2/\epsilon}$. The time step has been set to $\Delta t/\tau=5 \cdot 10^{-5}$, with $D_t \, \tau/\sigma^2 = k_B \, T$. The reduced temperature is set to $k_B T/\epsilon=0.1$ to reduce the role played by the thermal noise.
Note that, in our simulations, the rotational and translational diffusivities $D_r$ and $D_t$ are not coupled by the Einstein-Stokes relation $D_r =3 \, D_t/\sigma^2$, due to this relation is strictly true only at equilibrium. In this work, we consider three different shoulder heights of $\epsilon_s/\epsilon=1, 10$ and $100$, and the velocity propulsion of each active particle is $v_p \, \tau/\sigma =10$. Besides number density $\rho =N/L^2$ ranges from 0.01 up to about 1.


\subsection{Analysis tools: Detection of MIPS}

By implementing the Fortune's algorithm we compute the Voronoi tessellation, the area available to each particle and the local density, which has been traditionally used to detect the onset of MIPS. When this function presents just one mode, the particles form a one-phase system, whereas when the distribution is  clearly bimodal they are segregated to form a MIPS system, characterised by areas of high and low density. Recently, other tools have been used to identify MIPS in ABPs systems. For example, it has been shown that ABPs interacting via WCA\cite{WCA} potential  show non-Gaussian displacements at intermediate times. This can be detected by means of the following two-dimensional non-Gaussian parameter\cite{JOSE}. This parameter measure the deviation between the PDF with respect to Gaussian distribution, characteristic of pure Brownian motion. The non-Gaussian parameter is traditionally  used to describe anomalous and/or heterogeneous transport dynamics in equilibrium systems or approaching the glassy state.

\subsection{Analysis tools: Identifying complex structures}

Although the shoulder-type potential shown in Eq. \eqref{eq:shoulder_potential} has spherical symmetry, it has been shown to produce a rich variety of phases \cite{malescio2004stripe}. When the particles are active the system reaches states with density distributions showing more than two peaks. To identify each of the phases appearing in the system and to help detect MIPS, we have developed the next structure classification algorithm. As shown in Figs. \ref{fig:phase}(b)-(f), particles interacting via a shoulder-like potential present different types of local structures. The different local structures are characterized, in part, by being made up of a number of particles that have at least one neighbour closer than a certain fixed limit that connect the particles, $r_c=1.7\,\sigma$. This cutoff value, in the range between
the hard core and shoulder radii, has been carefully chosen so that particles whose shoulders overlap are considered as belonging to the same structure, while particles belonging to the second layer are not. After imposing the above criteria, we apply the following classification: 

\begin{itemize}

    \item \textit{Monomers} are structures composed by only one particle (see Fig. \ref{fig:phase}(b)).
    
    \item \textit{Gas-Like} phase is composed by a set of monomers with a local density, $\rho_0$, smaller than $\rho_c = 0.145 \sigma^2$, which corresponds to the close packing of particles with diameter $\sigma_s/\sigma=2.5$.

    \item \textit{Dimers} are structures composed by two connected particles (see Fig. \ref{fig:phase}(c)).
    
    \item \textit{Trimers} are structures composed of three particles, all with the other two particles at a distance of less than $r_c = 1.7 \sigma$ (see Fig. \ref{fig:phase}(d)).
    
    \item \textit{Chains} are structures formed by N particles ($N>2$) in which all particles are connected with less than 4 neighbours not connecting each other. Hence, chains can be very long and may have branches (Figure 1.e).
    
    \item Other complex structures are simply classified as \textit{dense phase}, and in them the constituent particles connected with more than three neighbours, or there are three neighbours connecting each other.
    
    In these structures, {\color{black} hexagonal order is detected by computing the hexagonal order parameter for each particle
\begin{equation}
\psi_6(k) = \frac{1}{n} \, \sum_{j\in N_k} \, \text{e}^{i \, 6 \, \theta_{kj}},
\label{eq:Psi6}
\end{equation}
where $\theta_{kj}$ is the angle between the vector $\vec{r}_{kj}$ and the x-axis, $N_k$ is the set of first Voronoi neighbors\cite{ginelli} of particle $k$. 
    }. Note that hexagonal order appears both in the diluted and  dense phases.
\end{itemize}

\subsection{Analysis tools: Computing the pressure within the system}

How to compute pressure in non-equilibrium systems is still under debate. In active particle suspensions,  some works\cite{mallory2014anomalous, takatori2014swim, solon2015pressure, speck2016ideal, ginot2015nonequilibrium} have proposed expressions analogous to state equations that include terms accounting for activity. In 2D ABPs systems the total pressure can be computed as\cite{sanoria2021influence}
\begin{equation}
    P= \frac{N k_B T}{A} + \frac{\gamma \, v_p}{2\, A \, D_r} \, \sum_{i=1}^N \left\langle \vec{n}_i \cdot \dot{\vec{r}}_i \right \rangle + \frac{1}{4A} \sum_{i=1}^N \sum_{j=1}^N  \vec{F}_{ij} \cdot \vec{r}_{ij},
    \label{eq:TotalPressure}
\end{equation}
where $\gamma=K_BT/D_t$ is the friction coefficient and the sum over the particles includes the contribution from the periodic boundary conditions \cite{winkler2015virial}.  $\dot{\vec{r}}_i$ can be computed by using Eq. \ref{eq:motion}, discarding the thermal noise, which has a small value compared to the other two terms. Note that the first term represents the ideal gas pressure, the second is the activity contribution, so called swimming pressure, $P_s$, and the last term includes the contribution of the interaction potential.  Hence, we can compute the local contribution of each particle i-th to the pressure as
\begin{equation}
    P^{(L)}_i=\frac{\gamma \, v_p}{2\, A \, D_r} \, \vec{n}_i \cdot \dot{\vec{r}}_i  + \frac{1}{4A} \sum_{j=1}^N  \vec{F}_{ij} \cdot \vec{r}_{ij},
    \label{eq:LocalPressure}
\end{equation}
the first term due to the activity and the second one including the contribution of the interaction potential.

\subsection{Analysis tools: Dynamical properties}

The relaxation dynamics of glasses is frequently examined by studying the decay of the self (incoherent) intermediate scattering function
\begin{equation}
    F(\vec{q},t) = \frac{1}{N} \sum_{j=1}^N \langle \exp\left[ i \vec{q} \cdot \left(\vec{r}_j(t)-\vec{r}_j(0)\right)\right]  \rangle.
    \label{eq:intermediate}
\end{equation}
Here, $\vec{r}_j(0)$ is the position of particle $j$ once the steady-state is reached (and time is set to 0) and $\vec{q}$ is the wave vector. Assuming that the stationary regime is isotropic and, thus, the intermediate scattering function only depends on the modulus of $q$ and time, we analyze the decay of $F(q^*,t)$, where $q^*$ is the smallest value of $q$ at which the static structural factor $S(q)$ shows a maximum. For a better understanding of the dynamics we also compute the Mean Square Displacement (MSD)
\begin{equation}
    MSD(t)=\left\langle  |\textbf{r}(t)-\textbf{r}(0)|^2  \right\rangle,
    \label{eq:MSD}
\end{equation}
as well as the long time effective diffusion coefficient $D_{eff}$, defined as the slope of the MSD at long times.\\


\noindent
\textcolor{black}{Inspired by Ref.\cite{caprini2020hidden,caprini2020spatial,caprini2020spontaneous} we have estimated the velocity fields when the system undergoes MIPS (See Supplementary Material for more details).
The velocity field have been computed as the displacement of the particles between two times $\vec{v}=\frac{\vec{x}(t+\tau)-\vec{x}(t)}{\tau}$, where $\tau$ is the unit of time.}

\section{Results}

In this work, we first study different structural properties of  ABPs suspensions characterised by various values of  the shoulder height. Next, we unravel their dynamical features.


 
 \subsection{Structural features}
 
\subsubsection{Soft ($\epsilon_s/\epsilon=1$) and hard ($\epsilon_s/\epsilon=100 $) shoulder  height}

\begin{figure}[h!]
     \includegraphics[width=0.45\textwidth]{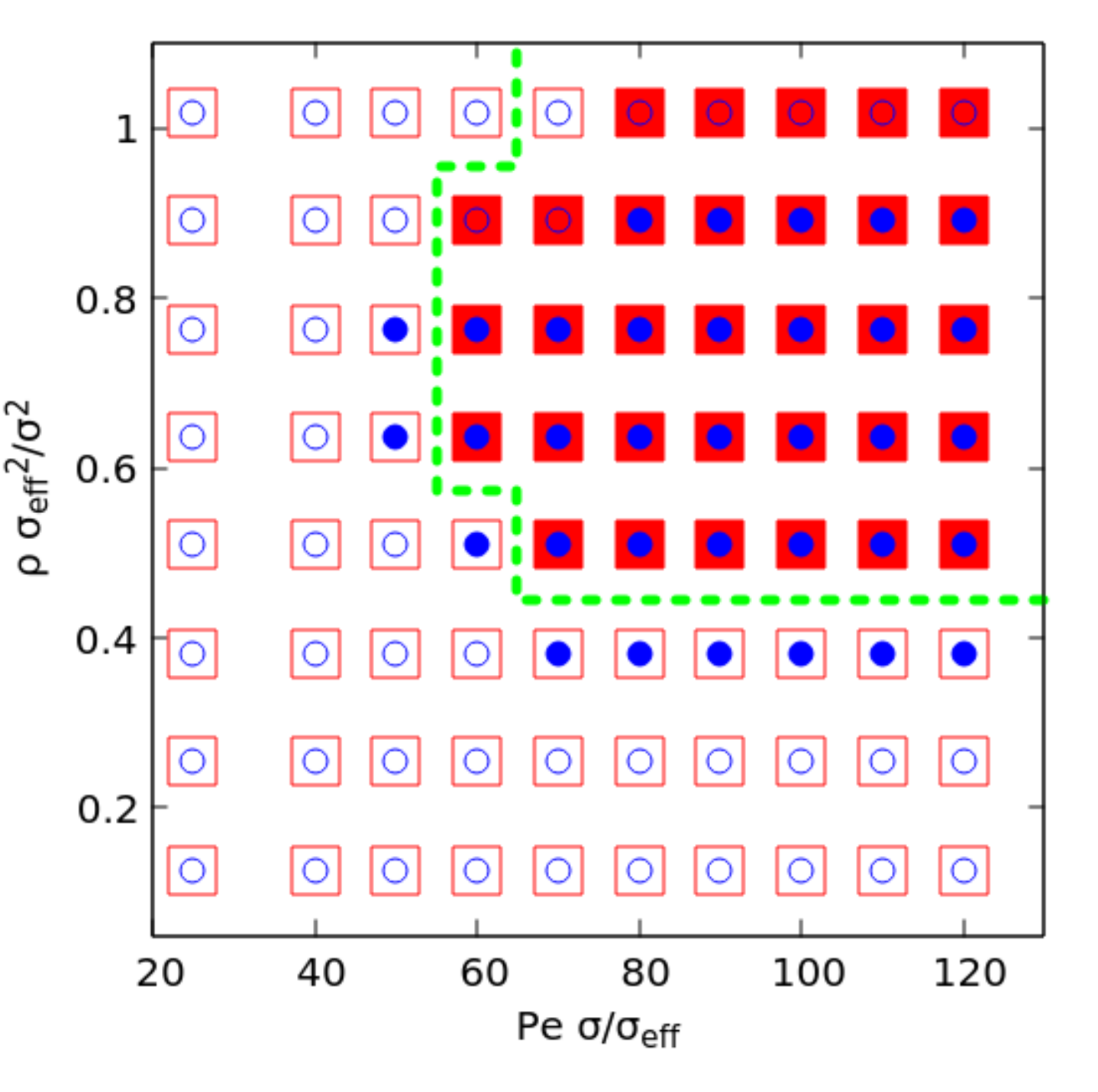}
    \caption{    
    Phase behaviour for a suspension of ABPs interacting through shoulder-like potential with $\epsilon_s/\epsilon=1$ (red squares) and $\epsilon_s/\epsilon=100$ (blue dots). Peclet  is modified by changing $D_r$, while keeping fixed $v_p=10$, $k_B T=0.1$ and $D_t=0.1$. The density and Pe axes were scaled ($\rho_{eff}=\sigma_s^2 \cdot \rho $ and Pe$_{eff}=\frac{Pe}{\sigma_s}$) so that both phase diagrams (for $\epsilon_s/\epsilon =1$ and 100) can be shown on the same plot. The empty symbols represent the system in an homogeneous phase and filled symbols in a MIPS phase. The green dashed line shows the onset of the MIPS regime when particles interact via a WCA potential.  }
    \label{fig:PhaseDiagram}
\end{figure}

To begin with, we explore the phase diagram of active particles interacting with a soft shoulder potential with height, $\epsilon_s/\epsilon =1$. In these conditions, active particles can easily overcome the shoulder of the potential generated by their neighbours and behave as if they have an effective diameter ${\sigma}_{eff} \approx \sigma$. Figure \ref{fig:PhaseDiagram} shows the $\rho$-Pe phase behaviour of such suspension (red squares). As in the case of WCA, for low $\rho$ and Pe the system is in a homogeneous state (empty squares), whereas when both density and Peclet number simultaneously increase, the system exhibits MIPS (filled squares). As expected, when the repulsive barrier height is low and can be easily overcome by the active particles, the phase diagram is not affected by the presence of a second length scale, being essentially the same as that of particles interacting through a WCA potential. Next, we explore the system when $\epsilon_s/\epsilon =100$. In this limit particles behave as if they have an effective diameter $\sigma_{eff} \approx \sigma_s$. Therefore, the phase diagram is similar to that exhibited when particles with diameter $\sigma_s$ interact via a WCA potential. In Figure \ref{fig:PhaseDiagram}, the different states depicted by the ABPs interacting through a shoulder potential with $\epsilon_s/\epsilon =100$ are shown with the filled or empty blue symbols. It is evident that both phase diagrams approximately coincide when the density and Pe number are properly scaled, except for two clear differences: i) The boundaries differ slightly at the lower limits of $\rho$ and Pe, due to differences existing between the slopes of the soft shell and the hard core potentials (see Fig. 1.a), and ii) at high density, the particles are allowed to overlap because even though the shoulder of the soft shell potential is rather high, the repulsion does not diverge to infinity. 

 
\begin{figure}[h!]
    \includegraphics[width=0.5\textwidth]{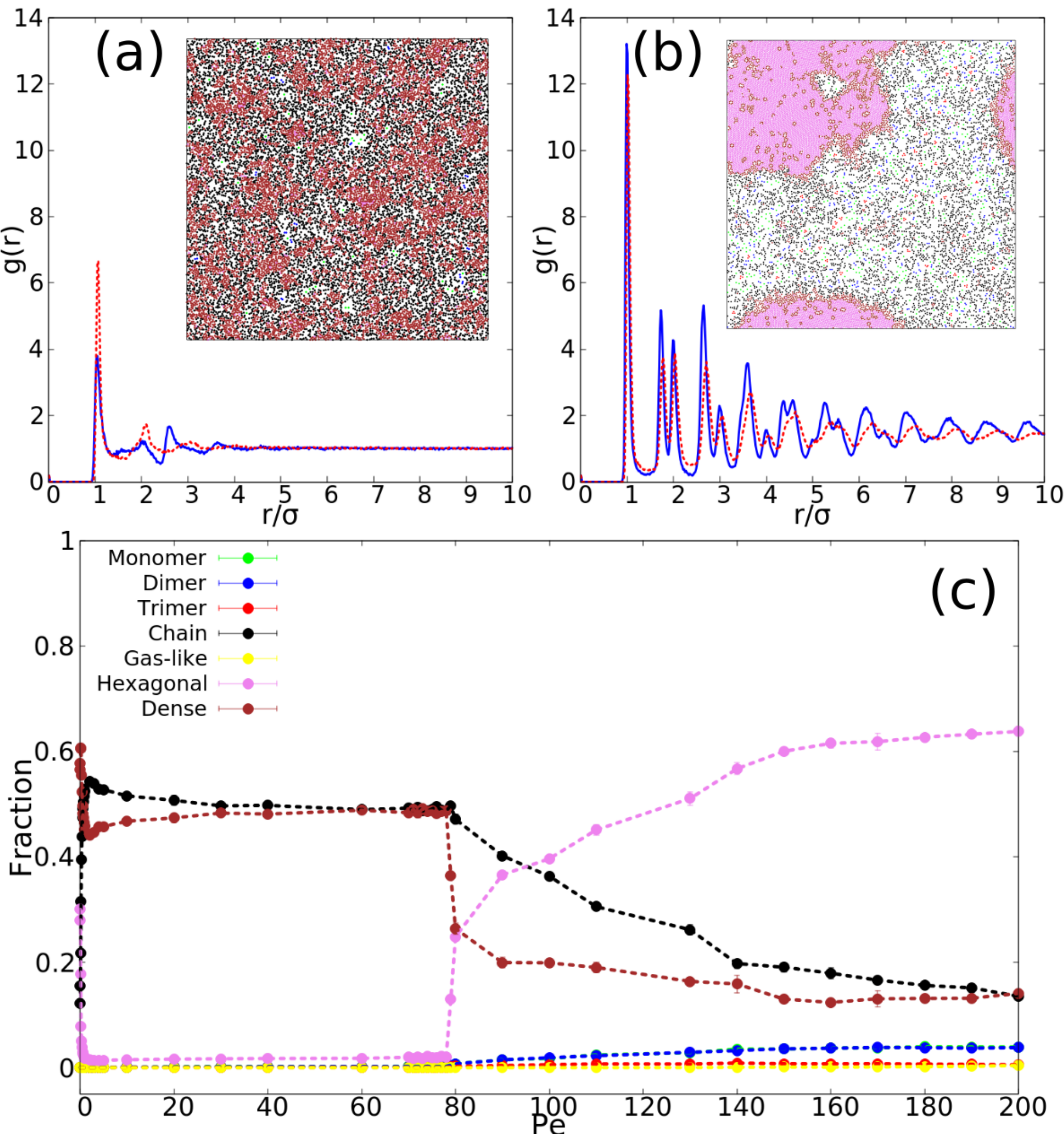}
    \caption{
    (a-b) Radial distribution functions for a system in the homogeneous (Pe= 50, a) and MIPS states (Pe= 120, b), when particles interact through the two lenghts scale repulsive potential with $\epsilon_s/\epsilon= 1$  (blue continuous line and corresponding snapshots) or the WCA potential (red dashed line). (c) Average percentage of monomers (green), dimers (blue), trimers (red), chains (black), high hexagonal order (violet), gas-like (yellow) and dense structures (brown) as a function of the Peclet for particles interacting via a shoulder-like potential with $\epsilon_s/\epsilon=1$.  In all calculations $\rho \, \sigma^2 = 0.509$.}
     \label{fig:gdr_001}
\end{figure}

In order to unravel whether the shape of the interaction potential plays a role in the structures formed, we analyse a global structural property such as the radial distribution function. In Fig. \ref{fig:gdr_001}a and b we show the radial distribution functions and the corresponding snapshots for a system characterised by  $\epsilon_s/\epsilon=1$  (in blue), at low  and high Peclet number (Pe=50 and 120, a and b, respectively). The results are compared to those obtained when ABPs interact through a WCA potential (dashed red line). In the homogeneous states, the radial distribution functions present two peaks when the ABPs interact via the shoulder potential, the first at the hard core distance and the second at the soft-shell distance, and only one when ABPs interact through the WCA potential. In the MIPS state (high Pe, Figure \ref{fig:gdr_001}.b) the $g(r)$ presents many peaks, which is characteristic of a highly ordered or crystalline structure. When comparing to  the system where ABPs interact via a WCA potential (dashed red lines in Figure \ref{fig:gdr_001}.a and b), we observe that the latter exhibits a similarly ordered structure. This supports the idea that the MIPS states in both systems are very similar. The visual inspection of the snapshots shows that, in the homogeneous state, most particles in the system are chain and dense structures. In the MIPS state mostly dimers, chains or trimers are found in the low density regions. The high density phases are comprised by different species. The interior of the high density phases consists of hexagonal lattice domains separated by boundaries. In the regions between the high and low density phases, particles arrange in chain-like structures.

In order to study how the structure distribution changes as a function of activity,  we focus on a system composed of ABPs interacting via the shoulder-like potential, at a given density ($\rho \, \sigma^2=0.509$) and quantify how the populations of different particles types change as a function of  activity. Figure \ref{fig:gdr_001}.c shows the average fraction of the different types of structures computed at different Peclet numbers.  At low Peclet number, when the system is in a homogeneous state, it mostly consists of dense and chain-like structures (as shown in the corresponding snapshot in Figure \ref{fig:gdr_001}.a).  At high Peclet number, we detect an increase in the percentage of structure with hexagonal order. When the system enters in the MIPS state (around Pe$\approx75$ at $\rho \, \sigma^2=0.509$), most of the chains become ordered and transform into hexagonal phase whereas the other phases change only slightly. For the sake of comparison, we have checked that, at the same conditions, when ABPs interact through a WCA potential {\color{black} particles only arrange in a} dense hexagonal state for larger values of the Peclet number.

\subsubsection{When  repulsion strength competes with  activity}

\begin{figure}[h!]
    \includegraphics[width=0.5\textwidth]{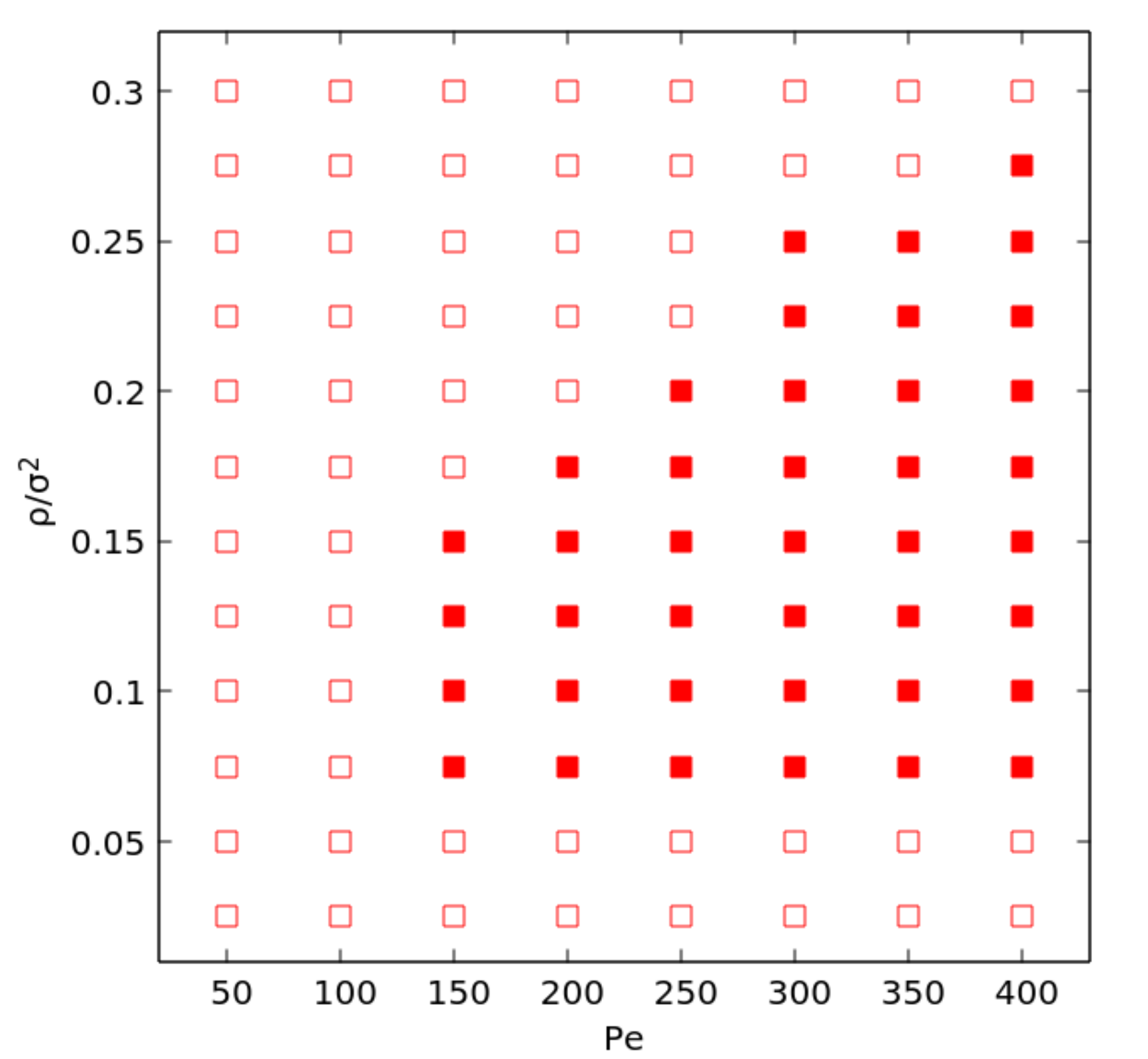} 
     \caption{ Phase behaviour for a suspension of ABPs interacting through a shoulder-like potential with $\epsilon/\epsilon_s=10$. Peclet  is varied  changing $D_r$, while keeping fixed $v_p=10$, $k_B T/\epsilon=0.1$ and $D_t \,  \tau^2/\sigma=0.1 $. Empty symbols represent the system in an homogeneous state and filled symbols in a MIPS state.}
    \label{fig:PhaseDiagram10}
\end{figure}

\begin{figure}[h!]
         \includegraphics[width=0.47\textwidth]{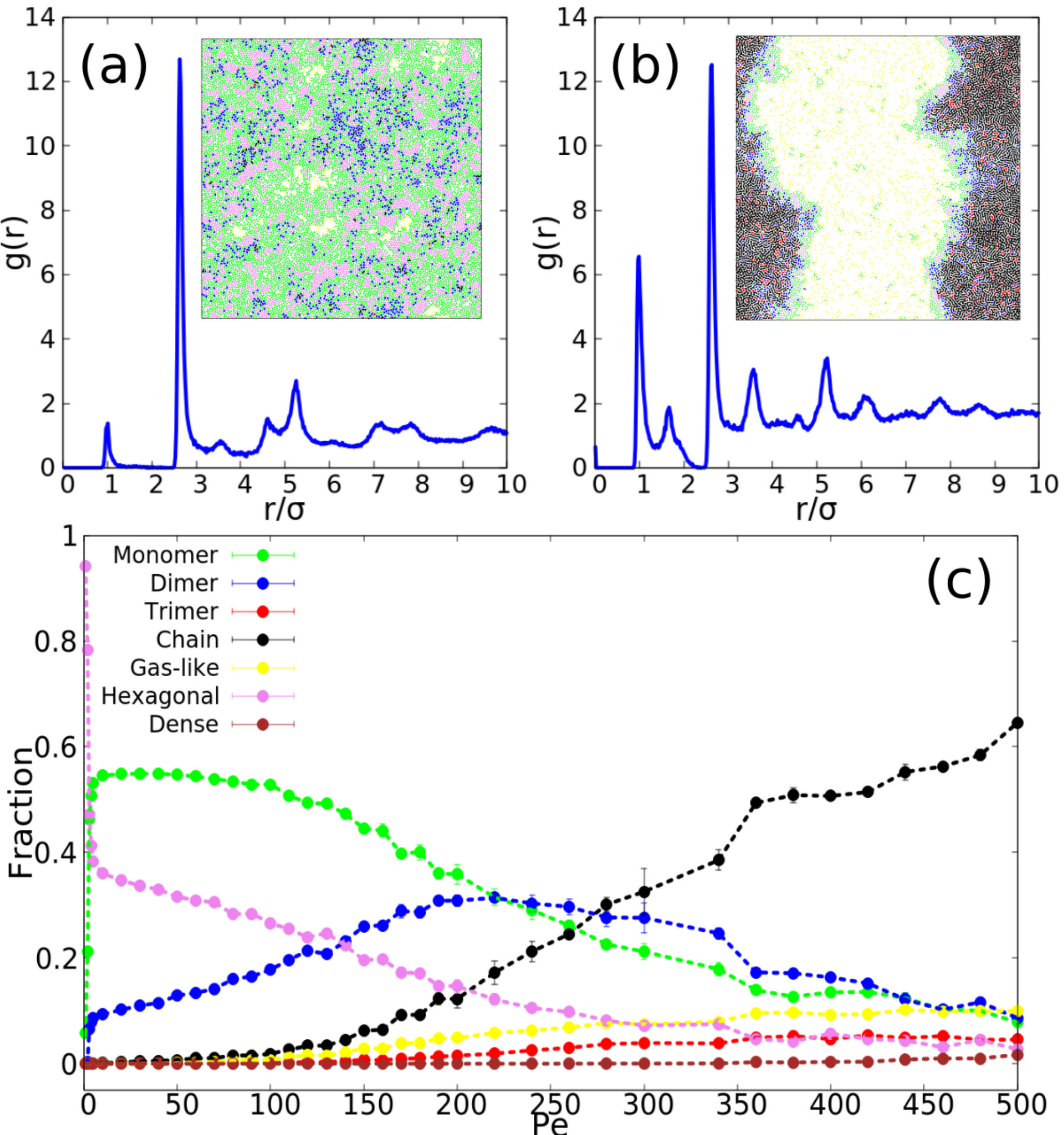}
      \caption{(a-b) Radial distribution function and corresponding snapshots for a system at $\rho \, \sigma^2=0.150$ in the homogeneous state (Pe$=80$, a) and  MIPS state (Pe$=500$, b); (c) Average percentage of monomers (in green), dimers (in blue), chains (in black), trimers (in red), high hexagonal order (violet), gas phase (yellow) and dense phase (purple) as a function of the Peclet number for ABPs interacting though a shoulder-like potential with $\epsilon_s/\epsilon=10$.}
     \label{fig:gdr_010}
\end{figure}

When $\epsilon_s/\epsilon =10$, the propulsion magnitude $v_p$ is not large enough for the particles to easily overcome the repulsive barrier of the corona as in the case when $\epsilon_s/\epsilon=1$. On the other hand, the shoulder barrier is  not high enough to prevent particle overlapping, as in the case when $\epsilon_s/\epsilon = 100$. Here, the soft repulsive shell starts playing an important role in the phase behaviour of the system. Figure \ref{fig:PhaseDiagram10} shows the $\rho$-Pe state diagram of such suspensions. Here, both length-scales are important and, therefore, it is not possible to scale the axes like in Fig. \ref{fig:PhaseDiagram} for the cases $\epsilon_s/\epsilon =1$ and $\epsilon_s/\epsilon =100$. 
As compared to the case when $\epsilon_s/\epsilon=1$ (Figure\ref{fig:PhaseDiagram}), {\color{black} here the MIPS  region shifts to lower densities and higher values of the Peclet number (compare axis ranges in Figures \ref{fig:PhaseDiagram} and \ref{fig:PhaseDiagram10})}. In the cases when $\epsilon_s/\epsilon=1$ and 100, MIPS is typically detected when the high density phase appears and the density distribution becomes bimodal. However, when $\epsilon_s/\epsilon=10$, the density distribution does not becomes clearly bimodal without showing clear signs of MIPS, so even at Pe=0 (so called passive case), particles organise in areas with different characteristic densities. Only at very high Pe, a very low density phase is created and MIPS can be clearly detected. Therefore, the mechanism of MIPS formation is completely different from the cases where only one length scale is relevant and MIPS state have perfect hexagonal order {\color{black}(see Supplementary Material for more details about passive and active systems)}. In this case, detection of MIPS boundary is slightly more complicated by means of the non-Gaussian parameter $\alpha_2$, using the method described in Ref \cite{JOSE}. In Fig. \ref{fig:gdr_010} we show the radial distribution functions (panels a and b), the corresponding snapshots  and the fraction of particles classified into the different structures (panel c) when $\epsilon_s/\epsilon=10$ and $\rho \, \sigma^2=0.150$,  slightly higher than the close-packing density $\rho_c/\sigma^2 \approx 0.145$. Figures 5 a and b clearly show: an homogeneous state at low Peclet number (Pe=80) and a MIPS state at high Peclet number (Pe=500).

In the homogeneous state (low Pe,  in Figure \ref{fig:gdr_010},a)  $g(r)$ presents several peaks, the highest one corresponding to  the soft-shell distance (2.5$\sigma$). The peak at the hard-core distance ($\sigma$) is due to the fraction of particles that form dimers (around 15\% of the particles in the system, see Fig. \ref{fig:gdr_010},c).  A visual inspection of a snapshot taken at the same conditions shows particles  mostly coloured in  green (monomers) or blue (dimers). On the other side, when the system is in a MIPS state (high Pe=500,  Figure \ref{fig:gdr_010} b) the peak at $r=\sigma$ increases significantly, which is a sign that many more particles have managed to jump over the shoulder of the potential. A visual inspection of a snapshot taken at the same state point {\color{black}(inset in Figure \ref{fig:gdr_010}.b)}, shows structures very different from the one presented in  Figure \ref{fig:gdr_001}.  While in the latter, the bulk of the dense phase mostly consisted of monomers surrounded by dimers, in the former the bulk of the dense phase consists of chains and a few dimers and trimers surrounded by a layer of monomers, some of them arranged according to a crystalline order. In Fig. \ref{fig:gdr_001}, we observe that the MIPS state shows a very well ordered hexagonal dense phase, with peaks in $g(r)$ that are periodically spaced, together with a high percentage of monomers arranged in hexagonal order. However, in Fig. \ref{fig:gdr_010} a and b, the peaks of $g(r)$ are more irregular, which is a sign of a more heterogeneous arrangement of particles inside the dense phase, with a significant percentage of particles forming chains, dimers and trimers, and a much smaller fraction of hexagonal monomer. At very high activity (Pe=500), chains and dimers become the most frequent structure in the system, followed by trimers and monomers, with fewer hexagonal structures only present at the boundaries of the dense phase (Fig. 5.c).

\begin{figure*}
    \centering
    \includegraphics[width=\textwidth]{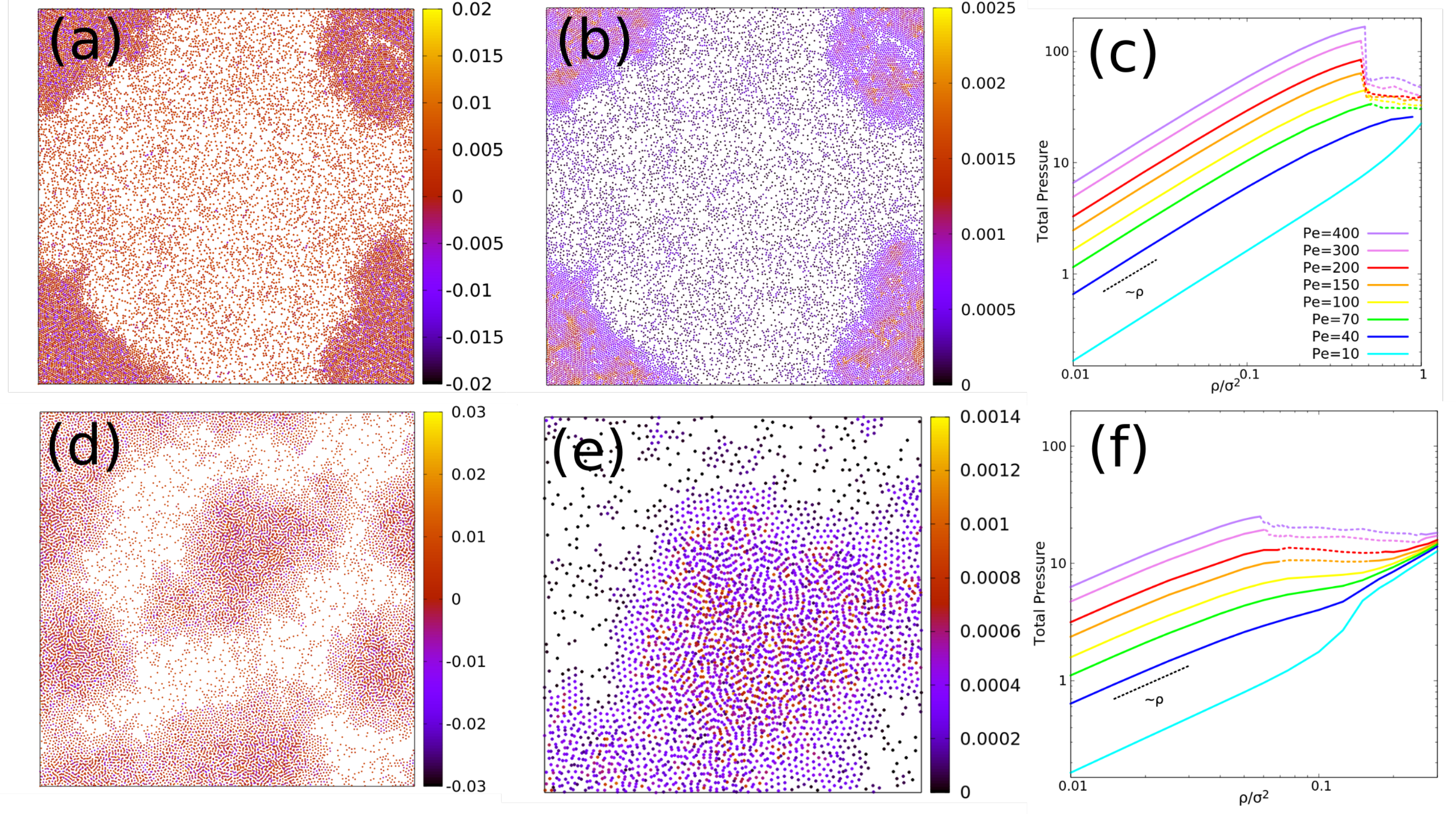}  
    \caption{Snapshots of the system for $\epsilon_s/\epsilon=1$, $\rho \sigma^2=0.509$ and Pe=100 (panels a-b) and $\epsilon_s/\epsilon=10$, $\rho \sigma^2= 0.150$, Pe=500 (panels d-e). Panels a and d show the swimming pressure, whereas panels b and e show interacting pressure. Note that the color bars are not in the same scale. {\color{black}For a better visualisation, panel e is a zoom-in of the system presented in panel d but with interacting pressure in the color code}. Panels c-f: Total Pressure vs Density for $\epsilon_s/\epsilon=1$ and $\epsilon_s/\epsilon=10$, respectively; different colours represent the value for Peclet number; continuous lines represent the density range where the system is in a homogeneous state, whereas dashed lines represent the density range where MIPS occurs. \label{fig:pressure}}
\end{figure*}

\noindent
\textcolor{black}{To conclude, we have explored the emergence of velocity correlations for three configurations that present MIPS with densities $\rho=0.1,0.2$ and 0.5, respectively. To better detect possible velocity correlations between particles within the dense phase, we have chosen to simulate these systems at very high activity (Pe=5000). Figures are presented in the Supplementary Material. In all cases studied, a clear alignment of the velocity vectors has been detected inside the dense phase, even when complex structures exist within. The same behaviour has been observed in a system of ABP interacting only via a one length scale repulsive potential \cite{caprini2020hidden,caprini2020spatial,caprini2020spontaneous}. }

\subsubsection{Pressure}

To better understand the structures formed within the separated phases reported in  Figures \ref{fig:gdr_001}  and \ref{fig:gdr_010}, we now compute both the global pressure and the local contribution to the pressure from each particle, Eq. \ref{eq:TotalPressure} and \ref{eq:LocalPressure}, respectively. Figure \ref{fig:pressure} shows the results obtained for two different cases, $\epsilon_s/\epsilon = 1$ (top) and $\epsilon_s/\epsilon = 10$ (bottom). Figures \ref{fig:pressure} a and b, show the snapshots of the system with $\epsilon_s/\epsilon=1.0$, $\rho \sigma^2=0.509$ and Pe=100, at conditions similar to the those imposed in Fig. \ref{fig:gdr_001}.b, coloured according to the value of the swim pressure and the interaction pressure, respectively. In parallel,  Fig. \ref{fig:gdr_001}.d and e show  snapshots of the system with $\epsilon_s/\epsilon=10.0$, $\rho \sigma^2=0.150$ and Pe=500 (conditions as those imposed in Fig. \ref{fig:gdr_010}b). Both snapshots are taken in the MIPS region (see the corresponding phase diagrams in Figures \ref{fig:PhaseDiagram} and \ref{fig:PhaseDiagram10}). Since the particles in the gas phase do not interact and their propulsion velocity is constant, the swimming pressure in the two exposed conditions (Figure \ref{fig:pressure}.a and d) is similar for all particles in the gas phase. In the areas inside the dense regions, interactive term in Eq. \ref{eq:LocalPressure} is much larger than those due to the particle activity and to the thermal noise. Hence from Eq. \ref{eq:motion} we can compute the velocity of the particle  $\dot{\vec{r}}_i \approx \frac{D_t}{k_BT} \vec{F}_i$, where $\vec{F}_i=\nabla V_i$ is the total conservative force acting over the particle. Here, we set $k_BT=0.1$ and $D_t=0.1$. Since in Eq. \ref{eq:LocalPressure}, the swimming pressure is given $P_s =\frac{\gamma \, v_p}{2\, A \, D_r} \, \vec{n}_i \cdot \dot{\vec{r}}_i $. In dense phase the swimming pressure is given by $P_s =\frac{\gamma \, v_p \, k_B T}{2\, A \, D_r\, D_t} \, \vec{n}_i \cdot \vec{F}_i $. Using this approximated Equation, the negative pressure observed in dense phase (Fig. \ref{fig:pressure}.d) is due to the events where $\vec{n}_i \cdot \vec{F}_i<0$, i.e. when the interactions and active force acting on the particles are anti-parallel. We postulate meanwhile the pure repulsive interaction force try to avoid the aggregation, the active force tents to keep the particles in the dense phase. In the low density region collision events are not often, then the value of the swimming pressure is higher than in the high density region. Figures \ref{fig:pressure} b and e represent the contribution from the interaction potential to the pressure for the two systems in a MIPS state. Even when we observe a high value of the pressure inside the dense phase, in panel b ($\epsilon_s/\epsilon=1$) we appreciate higher values. When $\epsilon/\epsilon_s=1$, particles in the dense phase are monomers interacting directly with the hard core as discussed above. On the other hand, when $\epsilon_s/\epsilon=10$ in the dense phase are chains already overlapped can move through the shoulder without any important contribution to the force. {\color{black}Looking at \ref{fig:pressure}.e (Zoom of panel \ref{fig:pressure}.d depicting the interacting pressure through the color code)}, particles with high interactive pressure in the dense region are monomers interacting with the chains via soft shell. Finally, Figures \ref{fig:pressure}.c and f show the global pressure for $\epsilon_s/\epsilon=1$ and $\epsilon_s/\epsilon=10$ as a function of the density in a range of Peclet number between 10 and 400. Here, the pressure is computed by means of Eq. \ref{eq:TotalPressure} in both systems. At low density the contribution from the interaction potential to the pressure, is almost negligible. If we consider those conditions where the noise is small enough, the velocity of the particle is given by $\dot{\vec{r}}_i \approx \frac{D_t}{k_BT} |F_a |\, \vec{n}_i $ and the contribution of the activity to the pressure can be approximated to $P_s\approx \frac{1}{6}\,  \text{Pe} \,|F_a| \, \sigma \,  \rho$. Hence, in these conditions the total pressure is proportional to the number density, $P \approx \left( k_B T + \frac{1}{6}\,  \text{Pe} \,|F_a| \, \sigma \right) \rho$, as confirmed in Figure \ref{fig:pressure}. On the other hand, in the high density region we observe two different behaviours. For $\epsilon_s/\epsilon=1$ the pressure decreases drastically in the MIPS region (dashed line), as was reported in Ref\cite{sanoria2021influence}. For $\epsilon_s/\epsilon=10$ the system is almost frozen and the curves collapse. Figure \ref{fig:pressure}.f shows that the pressure is monotonic almost for all Pe values, but have an inflexion point around $\rho_c$. For densities above $\rho_c$ and high enough activity, particles have access to the soft shell, so the contribution from the interaction potential to the pressure is smaller and then the total pressure decrees.





\subsection{Dynamical features}

As discussed earlier, suspensions characterised by $\epsilon_s/\epsilon=1$ and $\epsilon_s/\epsilon=100$ behave like those composed by ABPs interacting through a WCA potential, whose dynamical behaviour was already studied in a previous work\cite{JOSE}. In the last section of the study we will mainly focus on the dynamics of systems where the shoulder height ($\epsilon_s/\epsilon=10$) competes with the energy provided by activity.


\subsubsection{Long time diffusion}

\begin{figure}[h!]
    \centering
    \includegraphics[width=0.5\textwidth]{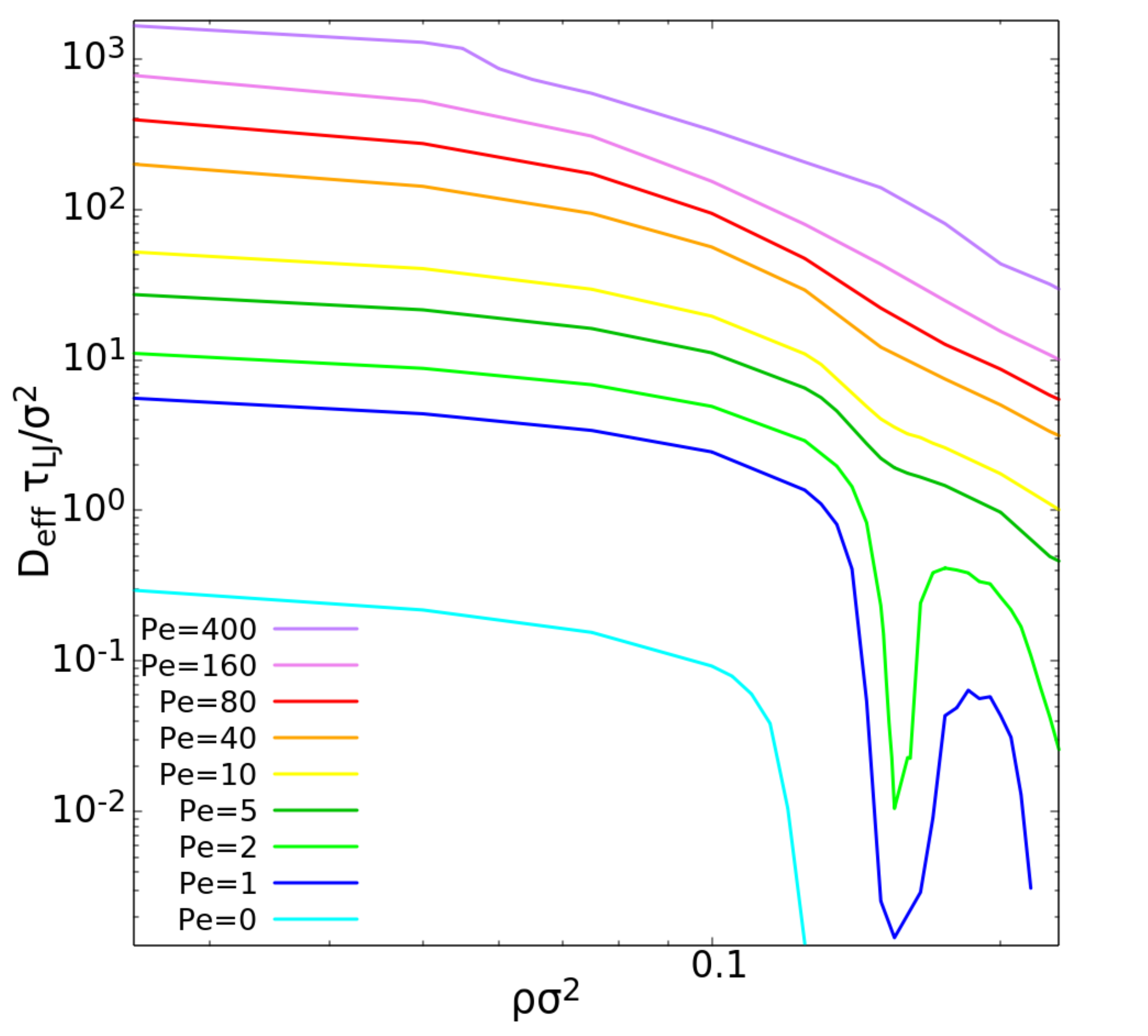}
     \caption{Diffusion coefficient versus density for $\epsilon_s/\epsilon=10$ and different values of the Peclet number, Pe=0, 1, 2, 5, 10, 40, 80, 160, 400.}
    \label{fig:Diffusion_010}
\end{figure}

In Figure \ref{fig:Diffusion_010} we report the diffusion coefficient computed from the long time behaviour of the Mean Square Displacement as a function of the system density and activity. We distinguish three different behaviours: one detected at Pe$=0$, one detected in the range between $0<$Pe$<5$ and another one for Pe$>5$. 
In the passive scenario, $D_{eff}$ monotonously decreases with density in the range between $0<\rho \, \sigma^2 < 0.150$. Beyond this value, the system is in a closed packed state: particles move only when they cross the region delimited by the shoulder of their neighbours and the only relevant length-scale is the width of shoulder. 

\noindent
When ($0<$Pe$<5$), the long time diffusion coefficient increases with activity in the range $0<\rho \, \sigma^2<0.15$, as expected. However, beyond the close-packing density, $D_{eff}$ increases again because the combination of mild activity and high density allows some particles to overcome the shoulders of their neighbours and move more easily. 

\noindent
At a scaled density slightly below 0.2, the diffusion constant goes through a maximum, whereas it decays back to zero for higher densities. When the activity is high (Pe$>5$), the repulsive shoulder becomes almost irrelevant and $D_{eff}$ decreases again monotonously over the whole density range, only dropping to zero at very high densities. The only sign of particles starting to overlap is the inflection point observed around $\rho \sigma^2=0.150$ which gets smoother as the activity increases, disappearing completely for Pe$>140$.



\noindent
In order to unravel the differences in dynamical behaviour between the passive and the active cases, we focus on two particular densities $\rho \, \sigma^2=0.150$ (just above the close packing density) and $\rho\, \sigma^2=0.20$ (33\% larger than the close packing density).  When particles are passive and $\rho\, \sigma^2=0.15$  particles are mostly monomers forming a hexagonal lattice, whereas at $\rho\, \sigma^2=0.20$ the system is disordered and particles form mostly dimers and chains (not showed here).  When Pe=1 and $\rho\, \sigma^2=0.150$, monomers still form an hexagonal crystal and the single particle diffusion coefficient barely changes.  However, at $\rho\, \sigma^2=0.20$, the inclusion of activity significantly changes the overall structures the chains disappear completely and dimers overall randomly distributed throughout the system. On the other hand, the diffusion coefficient increases by around 2 orders of magnitude with respect to the value measured in the passive system. At very high Peclet number (for example Pe$\approx$400), the behaviour detected at both densities is very similar, with signs of MIPS, analogous local structure inside the dense phase and comparable values of the diffusion coefficient. The only noticeable difference is the lower fraction of dilute phase at higher density. To summarize, at $\rho=0.150$ the system goes from stagnant at Pe<2 to very fast diffusion at Pe>2, with an increase of almost six orders of magnitude, while at $\rho \, \sigma^2=0.2$ monotonically changes from stagnant (at Pe=0) to very fast diffusion at large Pe.  

\noindent
{\color{black} A non monotonic behaviour of the diffusion coefficient has been already detected in suspensions of active glass formers 
(a binary mixture of WCA active  Ornstein Uhlenbeck particles). In Ref.\cite{berthier2017active}, the authors 
showed that the diffusion coefficient depended monotonically  on the persistence time for a given volume fraction, when the temperature was high. On the other side, the diffusion coefficient was not monotonous with persistence time at low temperature. A similar non monotonicity has been detected by  Ref.\cite{caprini2020diffusion} in an active  Ornstein Uhlenbeck particles suspension.}

\subsubsection{Self-Intermediate scattering function}

\begin{figure}
    \includegraphics[width=0.5\textwidth]{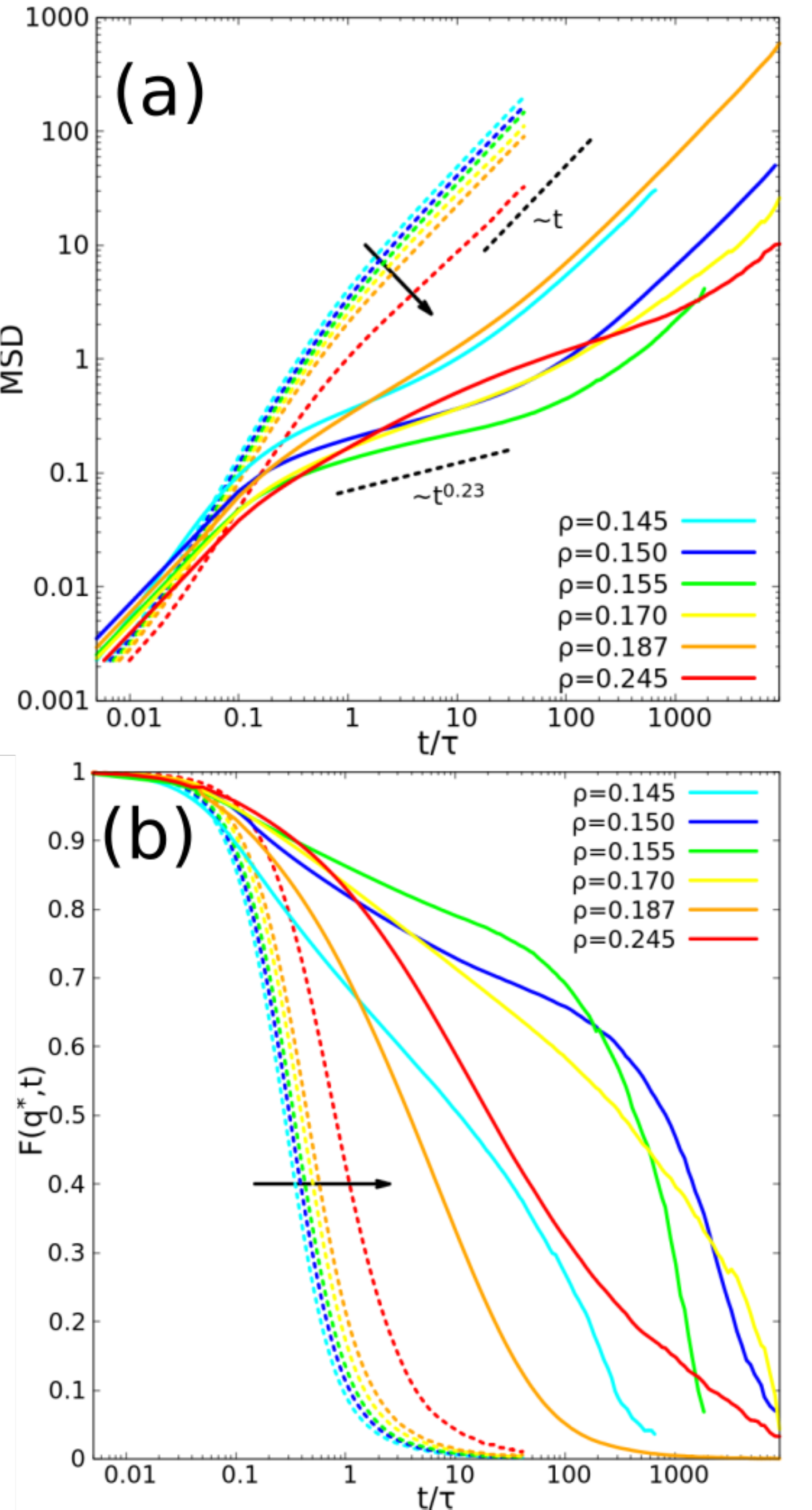}
    \caption{
    (a) Mean Square Displacement (MSD) for the system and (b) $F(q^*,t)$ vs simulation time for $\epsilon/\epsilon_s=10$ and two different values of Peclet number, Pe=1 (continuos line) and Pe=10 (dashed line). The densities used are showed in the legend.}
    \label{fig:slow_inter}
\end{figure}

The different qualitative dependence of the diffusion constant with respect to density observed in particles with low activity (Pe$<5$) and high activity (Pe$>5$) deserves a more detailed discussion. In this section, we explore the evolution of the mean-square displacement and the intermediate scattering function (Eq. 7) for systems with $\epsilon_s/\epsilon=10$, at different activities Pe=1 and 10, and at densities that range from $\rho \sigma^2=0.145$ (slightly below the overlap density $\rho_c$) to $\rho \sigma^2=0.245$ (well beyond the overlap density). Both results are shown in Fig. \ref{fig:slow_inter}. In Figure \ref{fig:slow_inter}.a, the MSD grows monotonically for both Pe=1 (continuous lines) and Pe=10 (dashed lines), but with qualitatively different behaviours. At Pe=10 the MSD shows a short-lived superdiffusive regime (MSD$\propto t^\alpha$, with $\alpha >1$) before moving to the terminal Fickian behaviour, while the MSD decreases monotonically with density, as expected from the results showed in Fig. \ref{fig:Diffusion_010}. Since the activity is large enough to overcome the effect of the soft shell, the dynamics is only determined by the hard core potential, and the system reproduces the behaviour depicted by solutions of ABPs interacting through a potential with a single relevant length-scale in the range between small to medium densities. At low activity (Pe=1, continuous lines) the MSD shows a completely different qualitative behaviour. First, it goes through a long-lived subdiffusive behaviour regime  ($\alpha<1$) before reaching the terminal Fickian regime. On the other hand, the dependence of the MSD with density is not monotonous. This rich diffusive behaviour is also reflected in the self intermediate scattering function, which is represented in Figure \ref{fig:slow_inter}b. For large activities (Pe=10) $F(q^*,t)$ decays almost as a single exponential, which is the characteristic behaviour of a particle diffusing in the Fickian regime. However, for small activity (Pe=1) it decrease much slower, with signs of the two-stage relaxation characteristic of glassy behaviour. In particular, at small densities ($\rho \, \sigma^2<0.187$) a two step decay is clearly observed. The emergence of the fast decay suggests that at these densities most particles diffuse along short distances or, equivalently, that they are trapped inside a cage. Only at long times, when the collective motion of particles opens the cage, particles can escape and $F(q^*,t)$ decays to zero. The terminal decay time, related to the long time diffusion coefficient, is again not monotonic, as expected from the results depicted in Fig. \ref{fig:Diffusion_010}. At higher densities ($\rho \, \sigma^2 =0.187$), the overlap of particles allows them to access the space of the soft shell and move more freely. That is reflected in the decay of $F(q^*,t)$, which now looks more like an stretched exponential, characteristic of a collection of diffusive particles that present a wide distribution of diffusion coefficients. Finally, when the density increases again ($\rho \, \sigma^2=0.245$), the system slows down again due to the increased density, as explained above, when discussing the mean-squared displacement.

As shown by Ref.\cite{mandal2020multiple}, an active glass former (ABP interacting via Kob Anderson) shows a thermal ageing when the persistence time is small. 
On the other side, when the persistence time is large,  ageing happens via two steps:  active athermal aging at short times and activity-driven aging at long times.

\section{Conclusions}

In this work, we study a system of Active Brownian Particles (ABPs) interacting through a continuous shoulder potential. Even for the passive case, a lot of interesting phenomena arise and particles self-organise in complex local structures (monomers, dimers, trimers, chains and others). One of the main points of this work is to characterise these structures and to study how the fraction of each one changes with the activity, focusing on the range of densities and activities where the system presents MIPS.

At constant propulsion velocity, we study three different values for the shoulder barrier strength, $\epsilon_s/\epsilon =$ 1, 10 and 100. Depending relative size of the activity sand the strength of the soft shell barrier, the particles can overlap the shoulders of their neighbours easily or not. In the cases $\epsilon_s/\epsilon =$ 1 and 100, just one the length-scale is relevant and, given that the shoulder potential is approximately equivalent to a WCA potential with an effective diameter, we can map the phase diagram for this two barrier values with the WCA one. We focus on the case $\epsilon_s/\epsilon=10$, when the soft shell and the hard-core are both relevant for the dynamics of the system. In this case the system also presents, for some values of density and Peclet number, a Motility Induce Phase Separation (MIPS) region characteristic of active systems but, in contrast with the results found for the single length-scale case, the structures found inside the MIPS state are not hexagonal. To better understand the case when the activity and the strength of the soft shell are similar, we explore the equation of state, and find that it is clearly different from the single length-scale scenario. If the potential only has one relevant length scale, the pressure shows a discontinuity when the system enters the MIPS state, due to the separation in low and high density regions. On the other hand, when the two length-scales play a relevant role in the system, this transition is much smoother.

Finally, we have explored the dynamical features of the system when both length-scales are relevant, by studying  the MSD and the self-intermediate scattering function. Our results show that, in the low activity regime, the topological cage produced by  the soft shell promotes non-monotonic increase of the diffusion coefficient with density. When this cage is efficient, {\color{black} the particles barely move} through the system. However, if the activity is strong enough, particles can overlap each other and have access to more space to diffuse easily.


Recently, Ref.\cite{klongvessa2019active,klongvessa2019nonmonotonic} have experimentally shown that  the approach to glass
transition in an active system can be mapped onto a passive supercooled liquid of soft colloids. 
However, the mapping failed  beyond the
glass transition, where  a non-monotonic response
of the relaxation time appeared.
The authors suggested that 
the  non-monotonicity could be caused by  two competing effects due to  activity: an extra energy causing cage breaking, and  propulsion's directionality  hindering cage exploration.
Therefore, they concluded that the
many-body nature of cooperative motion should be  taken
into account when dealing with active glasses.

On the one side, our  work demonstrates that active colloids characterised by two repulsive length scales show an anomalous   dynamics qualitatively similar to that observed in passive glasses characterised by two repulsive length scales. 
On the other side, since our results resemble those presented in \cite{klongvessa2019active,klongvessa2019nonmonotonic}, we might state that the conclusions drawn in \cite{klongvessa2019active,klongvessa2019nonmonotonic}  hold for  active particles characterised by either one or two repulsive length scales.
Therefore, the anomalous dynamics in a dense suspension of active Brownian particles interacting via shoulder potential is due to a synergy effect of   the two repulsive length scales in combination with  the  activity.


\section{Acknowledgements}

C. Valeriani and F. Martinez Pedrero acknowledge fundings from MINECO PID2019-105343GB-I00. C. Valeriani acknowledges fundings EUR2021-122001 from MINECO. J. Ramirez acknowledges fundings from MINECO PID2019-105898GA-C22.

\section{Data availability statement}
The data that support the findings of this study are available from the corresponding author upon reasonable request.


%






\end{document}